\documentstyle[aps,epsf,multicol,prl]{revtex}
\begin{document}
\title{Liquid-to-liquid phase transition in pancake vortex
systems  
}
\author{Joonhyun Yeo} 
\address{Department of Physics, Konkuk University,
Seoul, 143-701 Korea}
\author{M.\  A.\ Moore}
\address{Department of Physics, University of Manchester,
Manchester M13 9PL, United Kingdom} 
\date{\today}
\maketitle

\begin{abstract}
We study the thermodynamics of a model
of pancake vortices in layered superconductors. 
The model is based on the effective pair potential
for the pancake vortices derived from the
London approximation of a version of
the Lawrence-Doniach model which is valid for extreme type-II
superconductors.
Using the hypernetted-chain (HNC)
approximation,
we find that there is a temperature below which
multiple solutions to
the HNC  equations exist. By explicitly evaluating
the free energy for each
solution we find that the system undergoes a
first-order transition
between two vortex liquid phases. The low-temperature phase has
larger correlations along the field direction than the high-temperature phase.
We discuss the possible relation of this phase transition
to the liquid-to-liquid phase transition recently observed
in Y-Ba-Cu-O superconductors in high magnetic fields  in the
presence of disorder.
 
\end{abstract}
\pacs{PACS numbers: 74.20.De, 74.60.-w}
\begin{multicols}{2}
\narrowtext

\section{Introduction}

The properties of flux lines in high temperature
superconductors in a magnetic field have been under
theoretical and experimental
investigation for many years but surprises continue to appear.
The first-order transition line 
observed in experiments \cite{highm,highm2,lowm,lowm2} 
well below the upper critical field
$H_{c2} (T)$ is usually interpreted 
as a melting line 
of a vortex lattice into a vortex liquid.
For Y-Ba-Cu-O (YBCO) superconductors,
the first-order nature of the transition 
disappears at  critical points
in both high \cite{highm,highm2} 
and low\cite{lowm,lowm2} magnetic fields. 
Unlike the upper critical point, which is usually
attributed to the effect of disorder,
the value of the lower critical field  seems to be 
determined  by the oxygen content of the sample
\cite{lowm2}. In Ref.~\cite{km} a
different interpretation of the first order transition 
was given 
in terms of a transition 
between two liquid phases with different correlation 
lengths along the magnetic field direction. The disappearance
of the first-order transition line at low magnetic
fields is explained naturally
as a critical end-point in this picture.
For fields higher than the upper critical point,
the effect of disorder is increasingly important. 
In a recent experiment on YBCO \cite{highm2},
a rich phase diagram at fields above the
upper critical field was revealed, which included
two liquid phases separated by a 
second-order (continuous) transition. Although this transition
appears in the regime where the effect of disorder is large, it
is a reversible thermodynamic phase transition the nature of 
which is yet to be understood theoretically. We believe that the 
results of this paper will be a step towards the understanding of this 
liquid-to-liquid transition.

In Section II  we shall introduce a model
for the vortex system in 
a layered superconductor in a magnetic field perpendicular to 
the layers. We focus  for simplicity only on the case 
where the effect of disorder is negligible.
The model is based on
the London approximation of the Lawrence-Doniach model 
and consists of
a system of pancake vortices in each layer interacting via
an effective pair potential obtained by integrating out the degrees
of freedom other than those of the pancake vortices.
We find that the present model exhibits a first-order transition
between two liquid phases with very different correlations
along the field direction.
The state below the transition is very strongly correlated
along the field direction such that the pancake vortices
are likely to be stacked on top of each other rather as if in a stiff
vortex line, while above the transition in the high temperature phase
the vortex line is  dissociated
resulting in a phase with very short correlation lengths. 

We note that our liquid-to-liquid transition differs from 
another phase transition between two vortex liquid phases,
namely the disentanglement transition, that has been
studied and debated for some time  \cite{entangled}. 
In the present model, unlike below the disentanglement transition,
there is no long range order along the field direction.
On the other hand, it might be possible to associate 
the present liquid-to-liquid transition with the one observed
in the recent experiment \cite{highm2} on YBCO at fields above 
the upper critical point.  
Since the effect of disorder has not been considered in this model,
it is impossible to make a quantitative comparison of the present result
with the experimental results in Ref.~\onlinecite{highm2}.
Also the orders of the phase transition are different 
from each other. We observe, however, that 
a first order phase transition in the 
absence of disorder becomes
a continous transition when the disorder exceeds a critical
amount \cite{imry}. 
This will also be the case in our model. 
It is our contention that the 
basic mechanism that drives the liquid-to-liquid phase
transition in our model is also responsible for the 
transition observed in Ref.~\onlinecite{highm2}.

In this paper the hypernetted chain (HNC) approximation is used to
calculate thermodynamic
quantities and correlation functions.
The HNC approximation is often used 
for describing  correlations in 
classical liquids \cite{hm}. We set up and solve numerically
the HNC integral equations corresponding to the pair
potential for the pancake vortex system.
The first order transition emerges when
below a certain temperature the HNC equations admit {\it multiple} solutions. 
By calculating and comparing the free energy of these 
solutions we can determine the thermodynamically stable
branch of solutions. We believe it is quite rare in the HNC
analysis of classical liquids to find  multiple solutions
and first order transtions in this way.
Calculations on vortex liquids based on the HNC approximation
have also been done in Ref.~\onlinecite{cor}, but using
a pair potential which does {\it not} include the 
Josephson coupling between the layers as done  here.
The authors of Ref.~\cite{cor} used
only a phenomenological melting
criterion without considering the possibility of 
multiple solutions to their HNC equations.

The paper is organized as follows. In the next section we
introduce the model, derive the pair potential for the
pancake vortices, and set up the HNC equations. The results
based on the numerical solution to the HNC equations are presented 
and discussed in Section III.
 
\section{Model}

\subsection{Derivation of the pair potential}

Our starting point is 
the Lawrence-Doniach (LD) model \cite{ld}
for a layered
superconductor in a magnetic field  along the c-axis 
perpendicular to the layers.
With the order parameter in the n-th layer denoted by $\psi_n$,
the LD Hamiltonian is given by  
\begin{eqnarray}
&&H[\psi,\psi^*] = \sum_{n} d_0 \int d^2 {\bf r} 
\Big[ \alpha |\psi_n({\bf r})|^2
+\frac{\beta}{2}|\psi_n ({\bf r})|^4  \nonumber \\
&&~~~~~+\frac{1}{2m_{ab}}|(-i\hbar\nabla-\frac{e^*}{c}{\bf 
A})\psi_n|^2
\nonumber \\
&&~~~~~+\frac{\hbar^2}{2m_c d^2} |\psi_n ({\bf r}) - \psi_{n+1} 
({\bf r})|^2
\Big] , 
\end{eqnarray}
where $d_0$ is the layer thickness, $d$ the layer spacing and 
$\alpha$, $\beta$, $m_{ab}$, and $m_c$ phenomenological 
parameters.

In the London approximation the magnitude $|\psi_n|$
of the order parameter $\psi_n ({\bf r})= |\psi_n|\exp
(i \phi_n ({\bf r}))$ is held fixed and the fluctuations of 
only the phase $\phi_n ({\bf r})$ is considered. 
We consider the extreme type-II case where the magnetic
field ${\bf B}=\nabla\times{\bf A}$ is constant and uniform. 
The Hamiltonian becomes
after neglecting constant terms
\begin{equation}
{\cal H}=\sum_n \int d^2 {\bf r} \Big[ \frac{J}{2}\big(\nabla
\phi_n - {\bf a}\big)^2 + \frac{J}{2\gamma^2 d^2}  \big( \phi_{n+1}
-\phi_n \big)^2 \Big],
\label{hld1}
\end{equation}
where $J=\hbar^2 |\psi|^2 d_0 /m_{ab}$, 
${\bf a}= (2\pi / \Phi_0 )
{\bf A}$ with $\Phi_0 = h c /2e$,
and $\gamma=(m_c /m_{ab})^{1/2}$ is the anisotropy parameter. 
In the above equation
we approximated the inter-layer Josephson coupling term as
$2(1-\cos(\phi_{n+1}-\phi_n ))\simeq (\phi_{n+1}-
\phi_n )^2$. It is convenient to 
introduce two dimensionless parameters
to specify the system. They are 
$\Gamma\equiv 2\pi \beta J$ and $\eta\equiv
(a/\gamma d)^2$ where $\beta=1/k_B T$ is the 
inverse temperature and $a=(\pi\sigma)^{-1/2}$ is the
inter-pancake vortex spacing
with $\sigma$ being the number density of the pancake vortices
for each layer that is in turn determined by the
magnetic field $B$.  In terms of the physical parameters, 
they are given by 
\begin{equation}
\Gamma=\frac{2}{k_B T}(\frac{\Phi_0}{4\pi \lambda} )^2 ,~~~
\eta=\frac{1}{\pi\gamma^2 d^2}\big(\frac{\Phi_0}
{B}\big), \label{twop}
\end{equation}
where $\lambda$ is the penetration depth parallel to the layers.
High temperatures thus correspond to  small values of $\Gamma$,
low temperatures to high values of $\Gamma$, and $\eta$ is a measure
of the coupling between layers. 

In order to focus on the vortex degrees of freedom,
we decompose the phase variable $\phi_n$ into the ``spin-wave part''
$\varphi_n$ and the vortex part $\Theta_n$ for each layer, $\phi_n
=\varphi_n + \Theta_n$, and integrate out the spin-wave part. 
To do this
it is convenient to introduce the Fourier 
transform:
\[
\varphi_n ({\bf r})=\frac{1}{dN_l}\sum_{q} 
\int \frac{d^2 {\bf k}}{(2\pi)^2}
\widetilde{\varphi}_q ({\bf k})\exp(iqnd + i{\bf k}\cdot {\bf r} ). 
\]
In this paper we consider a system which consists 
of a stack of $N_l$ layers,
on which we impose  periodic boundary conditions 
in the direction perpendicular to the layers;
$\varphi_{n+N_l}({\bf r})=\varphi_n ({\bf r})$.
where $q=2\pi j /N_l d$ and we use $N_l$ integer values of $j$ in
the sum such that the wavevector $q\in [-\pi/d ,\pi/d)$ belongs to 
the first Brillouin zone. $\widetilde{\Theta}_q ({\bf k})$ is defined
in the same way. For future use we introduce
the hatted quantity 
$\widehat{\varphi}_n \equiv (\sigma /d )\widetilde{\varphi}_n$
for each function.
It is straightforward to integrate over $\varphi_q$
to obtain the vortex part of the Hamitonian given by
\begin{eqnarray}
&&{\cal H}_{V} = \frac{J}{2}\Big[ \int d^2 {\bf r} \sum_n
\big(\nabla \Theta_n - {\bf a}\big)^2  
+\frac{1}{d N_l}\sum_q \\
&&\times\int\frac{d^2{\bf k}}{(2\pi)^2}  
\frac{\frac{2\eta}{a^2} (1-\cos qd)}{k^2+\frac{2\eta}{a^2} (1-\cos qd)}
k^2 \widetilde{\Theta}_{-q}(-{\bf k})\widetilde{\Theta}_q
({\bf k})\Big]. \nonumber
\end{eqnarray}

The above Hamiltonian represents pancake vortices on each layer
interacting through some pair potential 
$v_n ({\bf r})$ which can be read off from the above equation
as follows. 
The key point in deriving the pair potential for
the pancake vortices is to note that one can write
the pancake vortex
density $\rho_n ({\bf r})=\sum_i
\delta^{(2)} ({\bf r}_n-{\bf r}^{(i)}_n )$ as
\[ 
\rho_n ({\bf r})=|\frac{1}{2\pi }\nabla\times\nabla\Theta_n ({\bf r})|=
\frac{1}{2\pi }\nabla^2 \Theta^\prime_n ({\bf r}),
\]
where $\Theta^\prime_n ({\bf r})$ is given by 
$\partial_x \Theta_n = -\partial_y \Theta^\prime_n$,
$\partial_y \Theta_n = \partial_x \Theta^\prime_n $.
In the above expression
${\bf r}^{(i)}_n$ are the positions of the pancake vortices.
We also note that $(\nabla\Theta_n )^2 = (\nabla\Theta^\prime_n)^2$
and $-k^2 \widetilde{\Theta^\prime}_q ({\bf k})=
\widetilde{\rho}_q ({\bf k})$.
Finally to get the pair potential
$v_n ({\bf r})$ , we write the Hamiltonian
in terms of $\rho_n$ as follows:
\[
{\cal H}_{V}=\frac{1}{2}\sum_{n,n^\prime}\int d^2 {\bf r}
d^2 {\bf r}^\prime \; \rho_n({\bf r})v_{n-n^\prime}
({\bf r}-{\bf r}^\prime) \rho_{n^\prime}({\bf r}^\prime ).
\]
For the case where
there is no inter-layer coupling, $\eta=0$, we can see 
using the above procedure that
the pair potential behaves as $1/k^2$,
{\it i.e.} logarithmically in real space and that
${\cal H}_V$ describes $N_l$ 
independent copies of the
two-dimensional one-component plasma (OCP) where
charged particles interact with each other via a potential
which is the logarithm 
of their spatial separation, all 
in the presence of a neutralizing background charge. 
The long-ranged nature
of the interaction is a direct consequence of
neglecting the screening of the magnetic field. 
The neutralizing background is provided here by the applied field
represented by ${\bf a}$.
In the presence of the inter-layer coupling we can
generalize this analysis and we find that the pair potential,
$\widehat{w}_q ({\bf k})\equiv\beta\widehat{v}_q ({\bf k})$ 
is given by
\begin{equation}
\widehat{w}_q({\bf k}) = k^2_D \Big[ \frac{2}{k^2} - \frac{1}
{k^2 + \frac{2\eta}{a^2}\big(1-\cos qd \big)} \Big] ,
\label{pot}
\end{equation}
where $k^2_D = 2 \pi \sigma \Gamma$. In real space, the potential energy 
between two pancakes in the same layer is logarithmic and repulsive: between 
pancakes in different layers there is an attractive Coulomb-like interaction.
It is this attractive interaction which is the origin of the liquid-liquid
transition and a low-temperature phase where a  pancake tends to
stack on top of one in the layer below  to form a stiff vortex line
in order to take advantage of the forces of attraction between them.

\subsection{HNC equations}

We shall set up and solve the HNC equations associated with the
pair potential (\ref{pot}). 
First the direct correlation function $c_n({\bf r})$ and
the pair correlation function $h_n ({\bf r})$ 
are related through
the Ornstein-Zernike relation,
\begin{equation}
\widehat{h}_q ({\bf k})=\frac{\widehat{c}_q ({\bf k})}
{1-\widehat{c}_q ({\bf k})} .\label{hnc1}
\end{equation}
The pair distribution function $g_n({\bf r})
=h_n ({\bf r})+1$ is given  
in terms of the other correlation functions via the HNC closure 
equation \cite{hm}, 
\begin{equation}
g_n({\bf r})=\exp\big[ -\beta v_n ({\bf r}) - c_n ({\bf r}) 
+ h_n({\bf r})
\big]. \label{hnc0}
\end{equation}

In the following we shall solve these equations numerically.
It is well known from the previous studies \cite{2docp} 
of the two-dimensional OCP 
using the HNC approximation, one should take special
care of the long-ranged logarithmic potential. In the present 
model the long-range tails in the pair potential are given
by the terms singular as $1/k^2$ for small $k$ in 
$\widehat{w}_q ({\bf k})$. These tails behave as 
$(2- \delta_{q,0})k^2_D /k^2$.
We separate
the pair potential into the short-range part with
an arbitrary momentum cutoff $Q$ and the 
remaining long-range part:
$\widehat{w}_q ({\bf k})=\widehat{w}^{(s)}_q ({\bf k})
+\widehat{w}^{(l)}_q ({\bf k})$, where
\begin{eqnarray}
\widehat{w}^{(s)}_q ({\bf k})&=& (2-\delta_{q,0})
\frac{k^2_D}{k^2 + Q^2} \nonumber \\ 
&-& (1 - \delta_{q,0})
\frac{k^2_D}{k^2 + \frac{2 \eta}{a^2}\big(1-
\cos(qd)\big)} , \\
\widehat{w}^{(l)}_q ({\bf k})&=& (2-\delta_{q,0})
\frac{k^2_D Q^2}{k^2 \big( k^2 +Q^2 \big)} .
\end{eqnarray}
According to Eq.~(\ref{hnc0}), the long-range part of 
the pair potential must be cancelled by a long-range
part of the direct correlation function, {\it i.e.}
$c_n ({\bf r})= c^{(s)}_n ({\bf r})+c^{(l)}_n ({\bf r})$
where $c^{(l)}_n ({\bf r})=-w^{(l)}_n ({\bf r})$ such that
Eq.~(\ref{hnc0}) only involves short-range functions:
\begin{equation}
g_n ({\bf r}) =\exp\big[-w^{(s)}_n ({\bf r})
-c^{(s)}_n ({\bf r}) + h_n ({\bf r})\big]. \label{hnc2}
\end{equation}

Eqs.~(\ref{hnc1}) and (\ref{hnc2}) form a closed set
which we solve iteratively. There are many equivalent 
iteration methods. Here we start from some function
$c^{(s)}_n ({\bf r})$ and calculate 
$\widehat{N}^{(s)}_n ({\bf k})\equiv \widehat{h}_n ({\bf k})-
\widehat{c}^{(s)}_n ({\bf k})$ from Eq.~(\ref{hnc1}).
Then we obtain the updated $c^{(s)}_n ({\bf r})$ from
Eq.~(\ref{hnc2}) and the definition $N^{(s)}_n ({\bf r})$.
In the numerical calculation it is crucial to have 
rapid convergence. Here we employ the method introduced
in Ref.~\cite{ng}, where a combination of several previous
steps were used as an updated $c^{(s)}_n ({\bf r})$. 
We utilize the results from the 
{\it five} previous iterations to produce a new input
function.
For small values of $\Gamma$ it is sufficient to start
from $c^{(s)}_n ({\bf r})=0$ as the initial function. We then solve
the HNC equations for fixed $\eta$ and
for increasing values of $\Gamma$ using
the solution obtained for lower values of $\Gamma$.  

\section{Results and Discussion}

In this section we present the numerical solutions to the HNC equations
described above. Our calculation is done mainly on a system
consisting of $N_l=11$ layers. In the last part of the section
we consider how the solutions behave for larger values of $N_l$.

One of the most important results of the present
work is that
the HNC equations describing a pancake vortex liquid admit
multiple solutions over a range of values of  $\eta$ and $\Gamma$. This
arises in the numerical calculation in the following manner.
We find that the solutions which are obtained in the way described
above cease to exist
above some value $\Gamma_I$. We call these solutions 
type-I. For $\Gamma$ smaller than $\Gamma_I$ we were able to find 
another type of solutions (type II) which are different
from the type-I solutions but which approaches the type-I solution
as $\Gamma\rightarrow\Gamma_I$. The type-II solutions continue
to exist as long as $\Gamma\geq\Gamma_{II}$ for some $\Gamma_{II}$. 
Below $\Gamma_{II}$
again the type-II solution disappears. 
We then find yet another solution (named type-III) 
for $\Gamma\geq\Gamma_{II}$ which are different from the type-II and type-I
solutions. 

Among the three solutions that exist for 
$\Gamma_{II}\leq\Gamma\leq\Gamma_I$,
only one will be a true thermodynamically stable solution that
describes the system in equilibrium. 
The solution which is stable can be determined by finding which of the three
solutions has the lowest
free energy for given $\Gamma$.
The free energy can be
evaluated using the correlation functions calculated 
within the HNC approximation \cite{morita}. Explicitly
we calculate the excess (over the ideal gas) free energy  
$F^{\rm ex}$,
or the dimensionless free energy per particle $f^{\rm ex}\equiv
F^{\rm ex} /2\pi JN $ with $N$ being the total number of
pancakes in the system as follows: \cite{morita}
\begin{eqnarray} 
&&f^{\rm ex} = \frac{\sigma}{2\Gamma}\sum_n\int
d^2 {\bf r} \Big[ -c^{(s)}_n ({\bf r}) 
-w^{(s)}_n ({\bf r}) \nonumber \\
&&~~~~~~~~~~~~~~+ \frac{1}{2} h^2_n ({\bf r})
-h_n ({\bf r}) c_n ({\bf r}) \Big]  \\
&&~~~+\frac{1}{2\sigma N_l\Gamma}\sum_q \int\frac{d^2 {\bf k}}
{(2\pi)^2} \Big[-\log \big( 1+ \widehat{h}_q ({\bf k})
\big) + \widehat{h}_q ({\bf k})\Big] . \nonumber
\end{eqnarray}
Figure 1 shows a typical plot of the free energy as a function of 
$\Gamma$ for fixed $\eta$. We see that the 
type-II solutions
are everywhere unstable and that
$\Gamma_I$ and $\Gamma_{II}$ are two spinodal points. 
We can identify $\Gamma_c$ where the first order phase transition
occurs such that above and below this value the system is
described by the type-III and type-I solutions respectively. 

It is important to note that
this first-order transition is not a melting transition
nor does it involve the existence of  long-range order
in the direction perpendicular to the layers.
This can be most easily seen from the structure factor
defined by 
\[
S_n ({\bf k})=1 + \frac{1}{dN_l }\sum_q \exp(iqnd) \widehat{h}_q
({\bf k}).
\] 
Above and below the transition temperature $\Gamma_c$ the in-plane 
({\it i.e.} $n=0$) correlation is almost the same as can be seen from 
the structure factor in $k$-space in Fig.~2 (a) and also from the 
pair correlation function $h_0 (r)$ in real space Fig.~3 (a). 
We also note that no crystalline long range order exists
below the transition, since the peaks  have
finite widths. 
The most dramatic changes are observed 
in the inter-layer correlation. This can be seen by comparing
$S_n ({\bf k})$ for $n\neq 0$ above and below the 
transition as in Fig.~2 (b)-(d). Compared to the situation above
the transition, the system below the transition remains 
highly correlated even when the layers are far apart.
As one lowers the temperature (increases $\Gamma$) across the transition,
the pancake vortices
that were more or less independent of 
those in neighboring layers become instead highly correlated.

\begin{figure}
\centerline{\epsfxsize=7cm\epsfbox{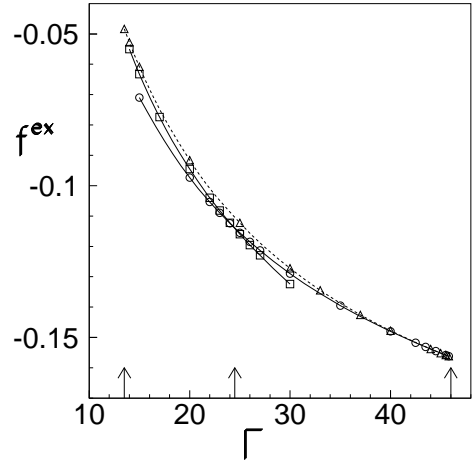}}
\vspace{10pt}
\caption{The excess free energe per pancake $f^{ex}$ 
as a function of $\Gamma$ for
$\eta=0.5$. The free energy for the type-I (circles) and 
type-III (squares) solutions 
are shown together with the unstable type-II
solutions (triangles, dotted line).
The arrow in the middle indicates the
transition temperature $\Gamma_c =24.5$. 
The arrows in the left and right of $\Gamma_c$ 
represent the spinodal points $\Gamma_{II} \simeq 13.0$ and
$\Gamma_I \simeq 49.0$,
respectively. 
}
\end{figure} 

\begin{figure}
\centerline{\epsfxsize=7cm\epsfbox{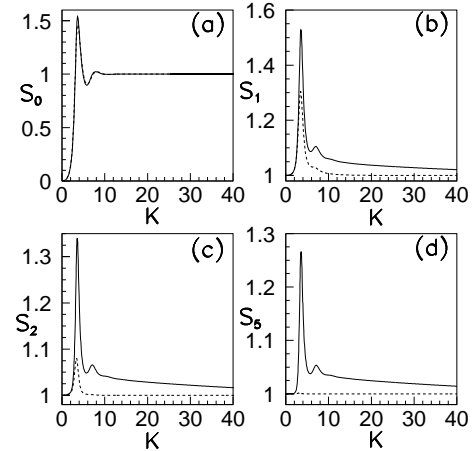}}
\vspace{10pt}
\caption{The structure factor $S_n (K)$ for (a) $n=0$, 
(b) $n=1$, (c) $n=2$ and (d) $n=5$ at $\eta=0.5$, where $K=ka$
is a dimensionless wavevector. 
The dotted lines are 
the type-I solutions at $\Gamma=24$ and the solid lines
are the type III solutions at $\Gamma=25$. 
}
\end{figure}

The phase below the transition described 
by the type-III solutions is characterized by
the tendency there of the pancake vortices  to stack
above each other along the field direction. 
This is clearly demonstrated by the behavior of  
the pair correlation function $h_n ({\bf r})$.
As can be seen 
from Fig.~3(c) the type-III solutions for $h_n (r)$ 
for $n\neq 0$ exhibit
a singular behavior for small $r$. This is in contrast to 
the type-I solutions shown in Fig.~3(b). The type-I
solutions are regular at $r=0$. The singular behavior
at small $r$ in type-III solutions
is also reflected in the very slow decay 
at large $k$ of the Fourier transformed quantities as seen in 
Fig.~2(b)-(d). 
We estimate that $h_n (r) \sim a_n / r^\alpha$ as $r\rightarrow 0$
where the constants $a_n$ and the non-universal exponent $\alpha$  depend 
on the temperature parameter $\Gamma$. From Eq.~(\ref{hnc0})
we can also deduce that the singularity in $h_n$ must be cancelled
by the same singularity in $c_n$, and this is confirmed in our numerical
calculation.
The singular behavior 
indicates that below the transition the pancake vortices in different layers 
have a large probability of being on top of each other
and so behave more like stiff vortex lines rather than individual
pancake vortices.
One way to see the correlation perpendicular to the layers is from 
the $n$-dependence of $h_n (r)$ for fixed $r$. For example 
for $r\simeq 0$, this quantity decreases
very rapidly to zero as seen from Fig.~3(b) for the type-I
solutions. This shows that the  correlation
length along the field direction is short. One the other hand, 
Fig.~3(d) shows that, below the transition,
$h_n(r)$  decays very slowly with $n$ and  
appears to approach a non-zero value suggesting that there might even be
long-range order  present.
As will be demonstrated later in this section, this is not the case.
In fact there seems to be no order parameter 
that describes  correlation across the layers which becomes
non-vanishing below the transition. Thus our transition seems to be best 
described as a liquid-liquid transition for which  the low-temperature 
phase has much longer correlations along the field direction than
the high-temperature phase. 

Using the calculated correlation functions, one can also determine 
other thermodynamic quantities like the excess internal energy
$U^{\rm ex}$. The dimensionless internal energy
per particle, 
$u^{\rm ex}=U^{\rm ex}/2\pi JN$
is given by
\[
u^{\rm ex}=\frac{\sigma}{2\Gamma}\sum_n\int d^2 
{\bf r} \; w_n ({\bf r})
h_n ({\bf r}).
\]
The specific heat per pancake excluding the ideal gas contribution
can be calculated by differentiating
the internal energy with respect to the temperature:
$c_v =N^{-1}d(U^{\rm ex} /dT) = k_B (du^{\rm ex}/d\Gamma^{-1})$.
The inset of Fig.~4 shows that 
the specific heat in our system obtained by
numerically differentiating $u^{\rm ex}$. The specific heat
is larger in the low temperature phase 
consistent with general features of an ordinary phase transition.
The liquid-to-liquid phase transition observed in 
\cite{highm2} is quite unusual in the sense that the 
low temperature phase has a smaller specific heat.
We believe this is also a consequence  of disorder
which in addition to changing the order of the transition might also modify 
the sign of the specific heat jump \cite{imry}.

\begin{figure}
\centerline{\epsfxsize=7cm\epsfbox{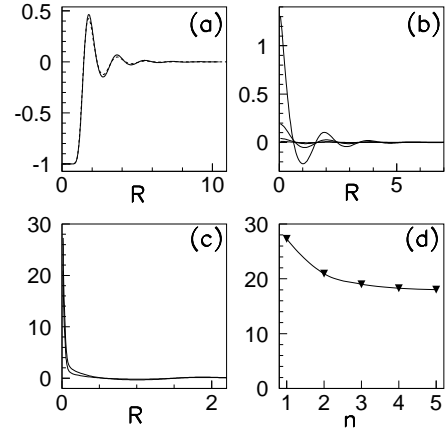}}
\vspace{10pt}
\caption{The pair correlation function $h_n (R)$ for $\eta=0.5$ 
from the type-I ($\Gamma=24$) and type-III ($\Gamma=25$)
solutions with dimensionless $R=r/a$. 
(a) $h_0 (R)$ from the type-I (dotted line)
and type-III (solid line) solutions; (b) The type-I $h_n (R)$
for $n=1,2,3$ and $4$.
The overall magnitude of $h_n (r)$ gets smaller
for larger $n$; (c) The type-III $h_n (R)$ for $n=1$ and $2$. 
(d) The type-III $h_n (r_0)$ as a function of $n$ 
for very small fixed $r_0\simeq 0.016$. 
}
\end{figure}

\begin{figure}
\centerline{\epsfxsize=7cm\epsfbox{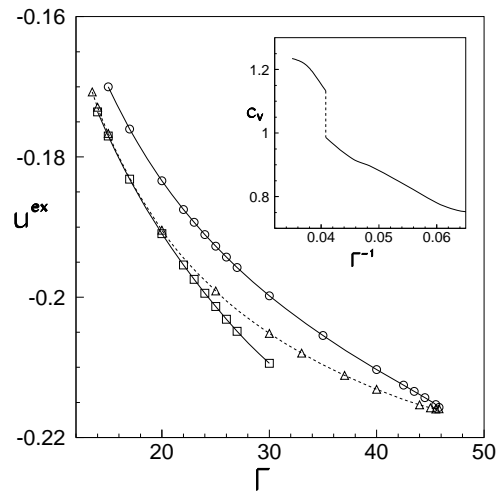}}
\vspace{10pt}
\caption{The excess internal energy $u^{\rm ex}$ per pancake as 
a function of $\Gamma$. The type-I (circles), type-II (triangles)
and type-III (squares) solutions are shown. The inset is 
the plot of the specific heat versus temperature $\Gamma^{-1}$.
}
\end{figure}

For other values of $\eta$, the first-order transition
point $\Gamma_c$ can in principle be determined
in the way  described above.
This will then
lead to a transition line in the physical $B$-$T$ plane
on using Eq.~(\ref{twop}).
Unfortunately, we found it quite difficult using our iterative technique 
to obtain type-III solutions starting from 
type-II solutions especially when $\eta$ becomes larger,
since they have a very different form 
from the type-II solutions especially in the small-$r$ region. 
In the present analysis
we were only able to identify a couple of transition points which
are shown in Fig.~5. 
It is, however, relatively easy to find the spinodal points
where the type-I and type-II solutions cease to exist.
The collection of these points up to
very large values $\eta$
are shown in Fig.~5. If there were a critical point
as in an ordinary liquid-gas phase transition, the two
spinodal lines would have to meet at the critical point.
Fig.~5 shows that the two spinodal lines almost run
parallel to each other in the large $\eta$ region.
Therefore within our model the phase transition
seems to continue into  both the large and small $\eta$ 
regime, implying that no critical points exist.
We also note that the slope of the 
phase transition line is opposite to those
of the spinodal lines. Since the transition line
should lie between two spinodal lines, we expect that 
as we go to the large-$\eta$ region the transition
line will change its direction and show a kind of 
re-entrant behavior in the $\eta$-$\Gamma$ space.

\begin{figure}
\centerline{\epsfxsize=7cm\epsfbox{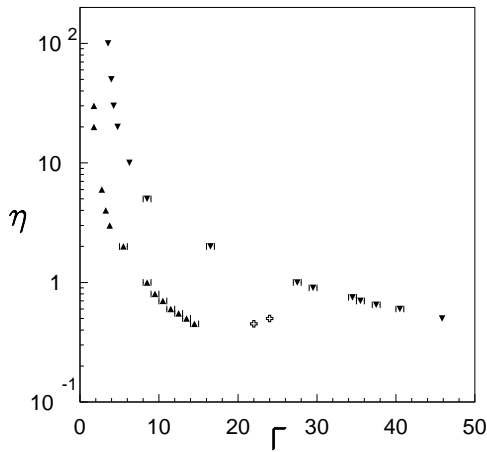}}
\vspace{10pt}
\caption{The spinodal points $\Gamma_I$ (inverted triangles)
and $\Gamma_{II}$ (triangles). The first-order transition 
points (crosses) for $\eta=0.45$ and $0.5$.
}
\end{figure}

The liquid phase described by the type-III solutions
is characterized by the inter-layer correlation functions 
becoming singular at small $r$ 
and decaying very slowly with $n$  (the layer separation)
at fixed $r$.  
For a system consisting of $N_l =11$ layers, 
Fig.~3(d) shows the correlation function almost 
becomes a non-zero constant for large layer separation $n$.
If this continues to be the case for larger systems,
then it would imply that
there is indeed long-range order below the transition in the
direction perpendicular 
to the layers. We therefore investigated 
the c-axis correlation function as done in Fig.~3(d) but for 
larger systems. The c-axis
correlation function shows a similar behavior for 
larger systems as  seen in Fig.~6. However, from Fig.~6 one
can also see that in the limit $N_l \rightarrow\infty$
the correlation function will decay to zero as $n$ becomes large,
implying that  there is no long-range order in the thermodynamic limit.
This analysis also tells us that in order to 
study the true thermodynamic behavior of the system, free from 
finite size effects, we will have
to consider a very large number of layers.
This will be a very challenging problem numerically.  

\begin{figure}
\centerline{\epsfxsize=7cm\epsfbox{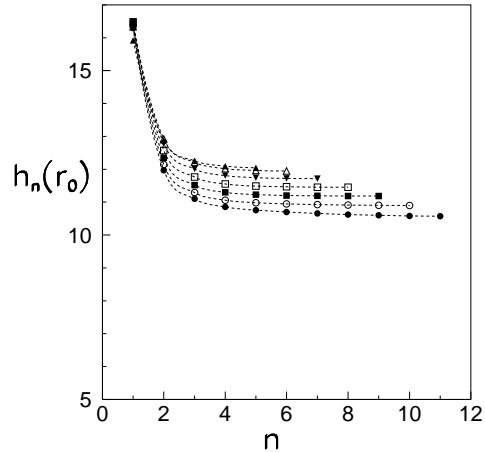}}
\vspace{10pt}
\caption{The pair correlation function $h_n (r_0)$
for very small fixed $r_0\simeq 0.016$ as functions of $n$.
The two parameters used are $\Gamma=17$ and $\eta=0.5$.
Systems consisting of 
$N_l =11, 13, \ldots, 23$ layers are presented. 
Because of the periodicity one only has to 
consider the values of $n=1,2, \ldots, (N_l -1)/2$.
}
\end{figure}

One aspect which we did not consider in this analysis
is the possibility of having crystalline solutions
of the HNC equations. In order to study this one has to 
start from a HNC functional which allows spatial
variation of the density. Eventually one has to evaluate 
and compare the free energies of 
the crystalline solution with the liquid state solutions.
Whether such solutions exist and have a lower free energy than the liquid
solution is left for future work. Other possibilities exist such as the
crystallization transition (if any) taking place at a lower temperature than 
the liquid-liquid transition. Evidence of two transitions has actually
been seen in a simulation of pancakes interacting with screened interactions
\cite{wj}. On this scenario one would argue that the presence of disorder
removes the crystalline phase at large fields leaving just the
liquid-liquid transition. In principle, because different symmetries are 
involved, a liquid-liquid transition followed by a liquid-crystal transition
on further cooling is a distinct possibility, although intuitively one might 
expect that the temperature interval between the liquid-liquid transition and 
the onset of crystallization might be small. Maybe only the presence of
disorder makes the liquid-liquid transition clearly visible as a distinct
transition.

In summary we have considered a model for pancake vortices in layered 
superconductors by deriving a pair potential for them. Using the HNC
approximation we have found that there is a first-order transition within 
the vortex liquid phase without any critical end-points. No long-range
order associated with the transition could be identified. However, in the 
low-temperature phase the pancake vortices stack up to form stiff vortex lines.
In order to study the rich phase structure exhibited by high temperature 
superconductors in a magnetic field it will be necessary to include the
effect of disorder into the present analysis. This will change our first
order transition into a continuous transition above some critical field 
\cite{imry}.

JY was supported 
by grant No.\ 1999-2-112-001-5 from the interdisciplinary
research program of KOSEF. MAM  would like to thank  F.\ Thalmann for
many useful discussions.

\end{multicols}

\end{document}